\documentclass[twocolumn,tighten,astrosymb,twocolappendix]{aastex701}

 \hypersetup{linkcolor=red,citecolor=blue,filecolor=cyan,urlcolor=purple}

\usepackage{ulem}
\renewcommand{\arraystretch}{1.1}
\usepackage{color}
\usepackage{array}
\usepackage{url}
\usepackage[bottom]{footmisc}
\usepackage{xcolor}
\usepackage{multirow}

\usepackage{graphicx}
\usepackage{multirow}
\usepackage{booktabs}
\usepackage{caption}
\usepackage{subcaption}
\usepackage{booktabs}
\usepackage{adjustbox}
\usepackage[toc,page]{appendix}
\usepackage{amsmath}
\usepackage{nccmath}
\usepackage{arydshln} 

\usepackage{aas_macros}
\usepackage{xspace}

\graphicspath{{./}{Figs/}}

\begin{document}

\title{JWST reveals a highly ionized nebula powered by an ultraluminous X-ray source \\ with a massive black hole in the metal-poor galaxy Holmberg II}

\author[orcid=0000-0001-5628-9120]{F.\,A.\,Fogantini}
\affiliation{Space Telescope Science Institute, 3700 San Martin Drive, Baltimore, MD 21218, USA}
\email[show]{ffogantini@stsci.edu}  

\author[orcid=0000-0002-6091-7924]{P.\,Zeidler}
\affiliation{AURA for ESA, Space Telescope Science Institute, 3700 San Martin Drive, Baltimore, MD 21218, USA}
\email{zeidler@stsci.edu}  

\author[orcid=0000-0003-0708-4414]{L.\,M.\,Oskinova}
\affiliation{Institut für Physik und Astronomie, Universität Potsdam, Karl-Liebknecht-Str. 24/25, 14476 Potsdam, Germany}\email{lida@astro.physik.uni-potsdam.de}  

\author[orcid=0000-0002-5082-5049]{T.\,Bao}
\affiliation{Institut für Physik und Astronomie, Universität Potsdam, Karl-Liebknecht-Str. 24/25, 14476 Potsdam, Germany}\email{xzbaotong@gmail.com}

\author[orcid=0000-0002-4970-6886]{W.-R.\,Hamann}
\affiliation{Institut für Physik und Astronomie, Universität Potsdam, Karl-Liebknecht-Str. 24/25, 14476 Potsdam, Germany}
\email{}

\author[orcid=0000-0002-0338-8181]{P.\,Marchant}
\affiliation{Sterrenkundig Observatorium, Universiteit Gent, Krijgslaan 281 S9
9000 Gent, Belgium}
\email{}

\author[orcid=0000-0001-8258-9813]{D.\,M.-A. Meyer}
\affiliation{Institute of Space Sciences (ICE, CSIC), Campus UAB, Carrer de Can Magrans s/n, 08193 Barcelona, Spain}
\email{meyer@ice.csic.es}

\author[orcid=0000-0002-5453-2788]{D.\,Pauli}
\affiliation{Institut für Physik und Astronomie, Universität Potsdam, Karl-Liebknecht-Str. 24/25, 14476 Potsdam, Germany}
\email{}

\author[orcid=0000-0002-3155-7237]{S.\,Reyero Serantes}
\affiliation{Centro de Astrobiología – Departamento de Astrofísica (CSIC-INTA), Camino Bajo del Castillo s/n, E-28692 Villanueva de la Cañada, Spain}
\email{}

\author[orcid=0000-0002-2090-9751]{A.\,A.\,C.\,Sander}
\affiliation{Zentrum für Astronomie der Universität Heidelberg, Astronomisches Rechen-Institut, Mönchhofstr. 12-14, 69120 Heidelberg, Germany}
\affiliation{Interdisziplinäres Zentrum für Wissenschaftliches Rechnen, Universität Heidelberg, Im Neuenheimer Feld 225, 69120 Heidelberg,
Germany}
\email{}

\author[orcid=0000-0003-0642-8107]{T.\,Shenar}
\affiliation{The School of Physics and Astronomy, Tel Aviv University, Tel Aviv 6997801, Israel}
\email[]{}

\author[orcid=0000-0002-4622-796X]{R.\,Soria}
\affiliation{INAF-Osservatorio Astrofisico di Torino, Strada Osservatorio 20, I-10025 Pino Torinese, Italy}
\affiliation{Sydney Institute for Astronomy, School of Physics, The University of Sydney, Sydney, NSW 2006, Australia}
\email{}

\begin{abstract}
Recent James Webb Space Telescope (JWST) observations have revealed fine-structure lines of ions with ionization potentials ${\gtrsim}50$\,eV, such as [Ne\,{\sc v}] and [O\,{\sc iv}], in nebular spectra of some low-metallicity galaxies, raising questions about the nature of the ionizing sources. Ultraluminous X-ray sources (ULXs) have been considered as potential contributors, but direct evidence linking individual ULXs to highly ionized nebulae is limited.
Here we present JWST IFU spectroscopy of the nebula surrounding the template ULX Ho\,II\,X--1 in the low-metallicity galaxy Holmberg\,II, already known to emit He\,{\sc ii}\,$\lambda4686$ and [O\,{\sc iv}]\,25.89\,$\mu$m lines. We resolve the infrared nebula on spatial scales of ${\sim}15-25$\,pc and report the first detections of [Ne\,{\sc v}] and [Ne\,{\sc vi}] in this source.
We find anisotropic ionizing conditions: emission from the highest-ionization species arises from compact regions offset from the ULX toward the previously detected radio knots, whereas lower-ionization species trace the wider nebula. 
The preferential alignment of the highest-ionization emission with the radio axis suggests anisotropic escape of the ionizing continuum and a physical connection with jet-ejection activity.
Spectral lines across the nebula are narrow and consistent with the instrumental resolution, indicating kinematically quiescent gas and disfavoring shocks as the dominant ionization mechanism. 
Together, these results provide direct evidence linking an individual ULX to a highly ionized nebula in a low-metallicity galaxy. These JWST observations of Ho~II~X--1 show that an accreting black hole in a low-metallicity galaxy is responsible for the extreme ionization of the surrounding nebula on ${\sim}100$\,pc scales.
\end{abstract}


\section{Introduction}
\label{sec:sec1}

\begin{figure*}
    \centering
    \includegraphics[width=0.525\linewidth]{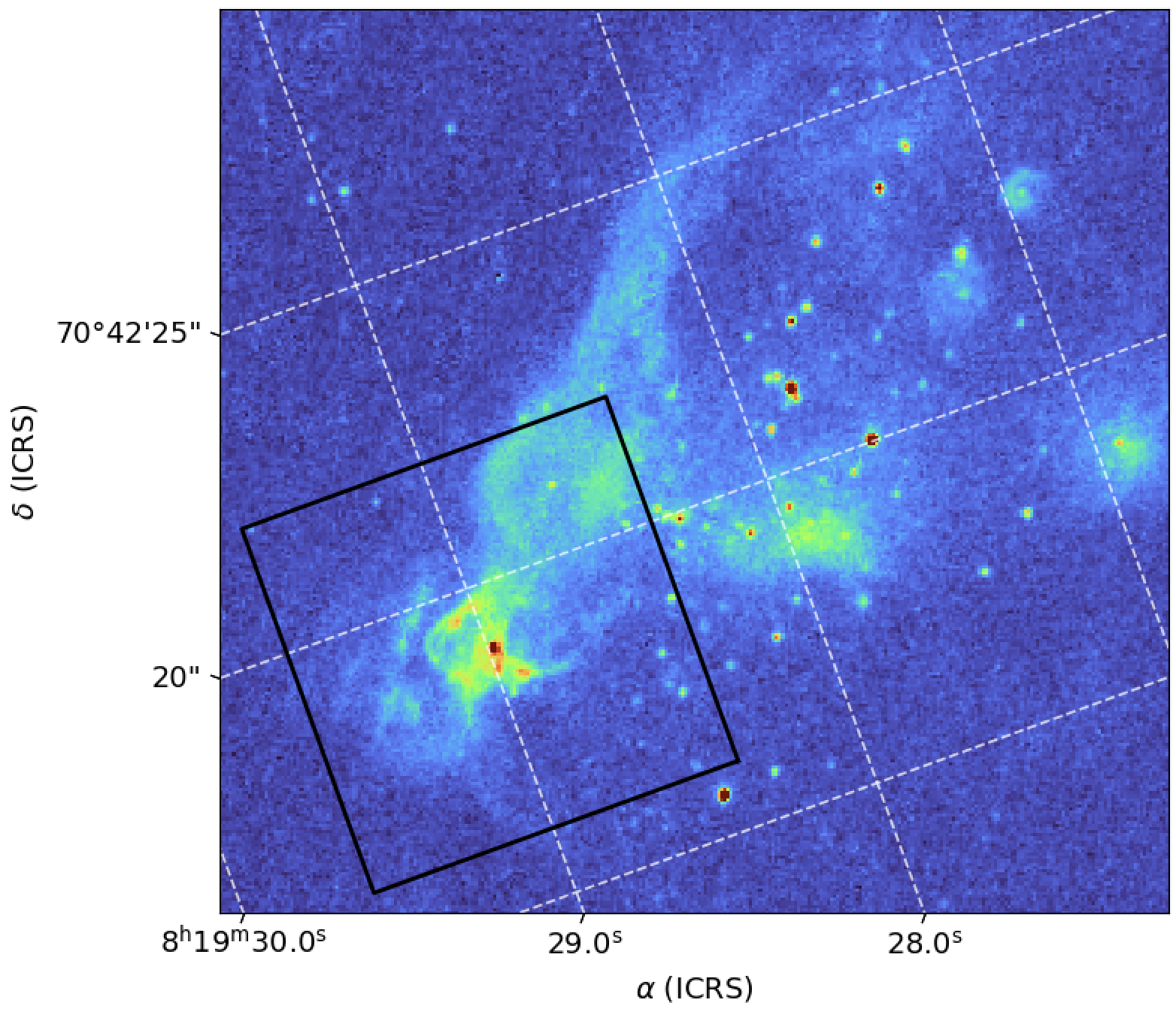}
    \includegraphics[width=0.465\linewidth]{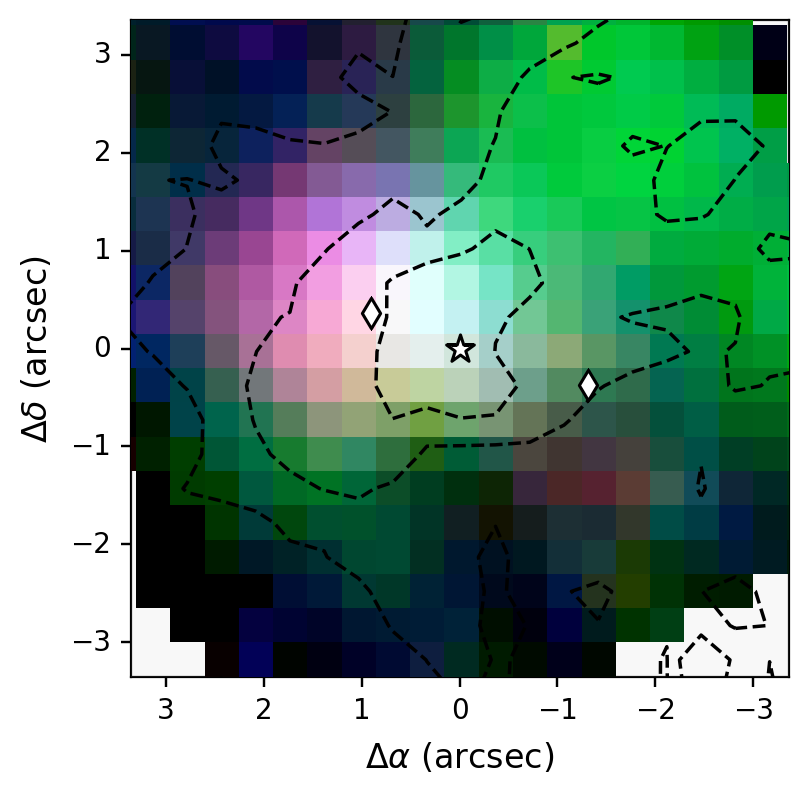}
    \caption{
    \textit{Left panel:} \textit{HST}/WFC F502N image of the Holmberg\,II nebula, covering approximately $15''{\times}15''$ ($1''{\simeq}16.5$\,pc at 3.4\,Mpc).  The black box ($6''\times6''$) is centred on the ULX Ho\,II\,X-1 and indicates the region displayed in the right panel. The color scale shows the relative F502N intensity in arbitrary units to illustrate the optical nebular morphology.\\
    \textit{Right panel:} \textit{JWST}/MIRI\,MRS RGB map of the Ho\,II\,X-1 nebula, expressed in offset celestial coordinates from the ULX position (referred to as the XS region using a star symbol). Green: the [S\,{\sc iii}]\,$\sim$18.7\,$\micron$ emission. Red: [Ne\,{\sc v}]\,$\sim$14.3\,$\micron$. Blue: [O\,{\sc iv}]\,$\sim$25.9\,$\micron$ emission. Each surface-brightness map was normalized to its peak intensity and PSF-matched to the MIRI\,MRS Channel\,4 spatial resolution, filtering out spaxels with SNR\,$<$\,3. 
    Black dashed contours show the \textit{HST}/WFC F502N surface brightness at the 95, 99 and 99.7 percentile levels emission of Ho\,II\,X--1, smoothed with a Gaussian filter of $0\farcs25$ width to suppress small-scale fluctuations. Diamond markers indicate the centroids of the NE and SW radio knots.
    }
    \label{fig:rgbmap}
\end{figure*}

Ultraluminous X-ray sources (ULXs) are X-ray binaries with luminosities $L_\mathrm{X}\,{\gtrsim}\,10^{39}$\,erg\,s$^{-1}$, exceeding the Eddington limit for a $10\,M_\odot$ accreting black hole \citep{Kaaret2017}. The majority of ULXs  are accreting  in strongly super-Eddington regime driving strong and ultrafast disk outflows \citep{Kosec2018, King2023}. 
An additional  ULX population could be powered by black holes with masses $> 10\,M_\odot$ accreting at the Eddington level. Such systems are expected to be particularly common in low-metallicity environments \citep{Fragos2013, Marchant2017}. 

It has been suggested that ULXs powered by black holes can provide the photons required to ionize He\,{\sc ii}, potentially explaining the presence of He\,{\sc iii} nebulae in some metal-poor galaxies \citep{Schaerer2019}. 
However, the contribution of ULXs to the He\,{\sc ii}-ionizing photon budget of metal-poor galaxies remains debated \citep{2026arXiv260505548A}.

The spectral energy distribution (SED) of ULXs is uncertain in the extreme ultraviolet (EUV)  that governs the production of ions with ionization potentials (IP) greater than ${\sim}54$\,eV \citep{Garofali2024,Nagele2026arXiv}. 
Some ULXs can also be exceptionally bright at ultraviolet wavelengths; \cite{Kaaret2010} directly detected an ultraluminous UV source associated with NGC\,6946/MF\,16.
Furthermore,  ULXs are typically found in crowded fields in remote galaxies, therefore establishing the clear links between ionized nebulae and individual ULXs is difficult. Therefore, a direct confirmation of high-ionization conditions in an ULX nebula is very much needed.

The ULX nebulae, which are usually observed in optical and radio, have  a wide range of excitation and kinematic properties. While some nebulae display shock-excited expanding shells with velocities of several hundred km\,s$^{-1}$, other nebulae are quiescent and their spectra are dominated by narrow lines consistent with photoionization \citep{Pakull2002}. 

Distinguishing between shock and photoionized nebulae is not easy and does not provide a unique diagnostic of the accretion regime operating in the X-ray source. Photoionization is always present, and may dominate in some locations or trace phases of the gas even in the shock heated nebulae. Shocks can be driven by jets or disk winds. Hence, a nebula which is predominantly photoionized does not necessarily imply the absence of mechanical feedback. Emission-line ratios and ionization structure provide important constraints on the dominant excitation mechanism, while spatially resolved spectroscopy adds the ability to connect local ionization conditions with the kinematics of the gas \citep{Abolmasov2007,Huang2026}.

The advance of \textit{JWST} allows for the first time spatially resolved IR spectroscopy of an ULX nebula.
Here we present the first \textit{JWST} NIRSpec IFU and MIRI\,MRS spectroscopy of the template ULX Holmberg\,II\,X--1 (Ho\,II\,X--1 hereafter). 
The paper is organized as follows: the object is introduced in Sec.\,\ref{sec:sec2}, the data processing and analysis are presented in Sec.\,\ref{sec:sec3} and \ref{sec:sec4} correspondingly. Section\,\ref{sec:sec5} presents the discussion, while the conclusions are drawn in Sec.\,\ref{sec:sec6}.

\begin{figure*}
    \centering
    \includegraphics[width=\textwidth]{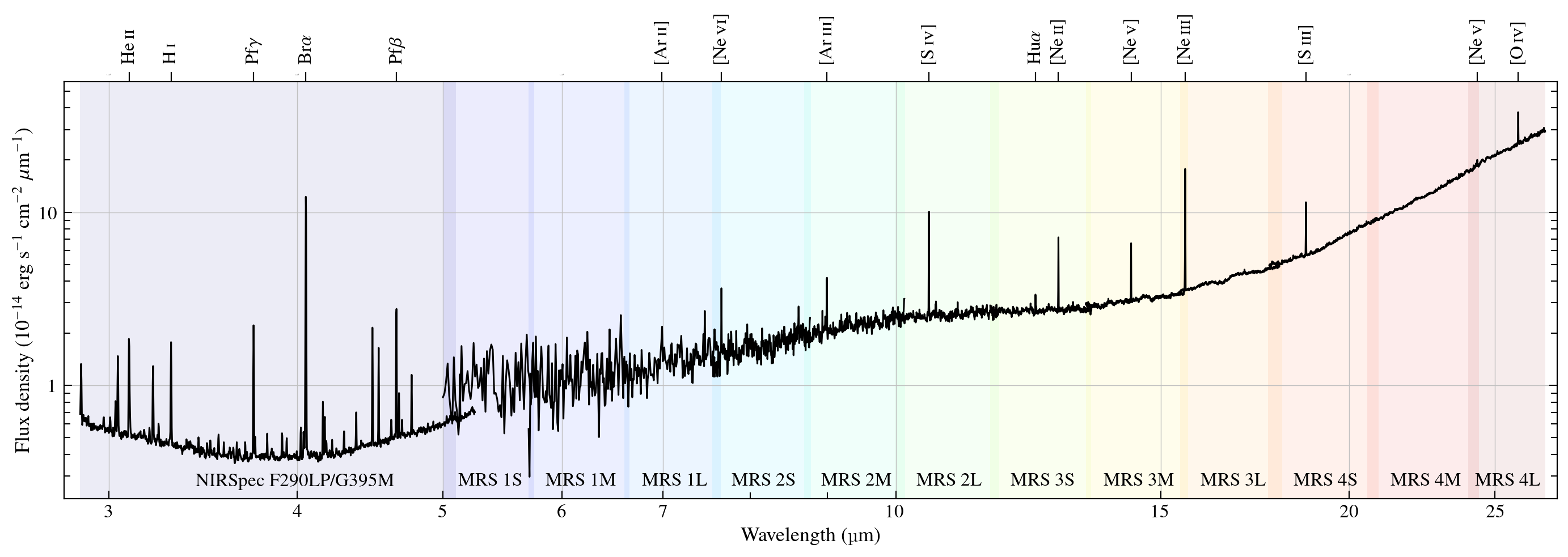}
    \caption{Broadband near- and mid-infrared spectrum of Ho\,II\,X--1 nebula derived from {\sc Stage~3} data cubes without background subtraction. The spectrum is extracted from a circular aperture with a radius of 1\arcsec centered on the ULX position (${\sim}16.5$\,pc) for display purposes only. The prominent emission lines are identified. MIRI MRS channel\,1 was rebinned linearly by a factor of 10, as this channel is the most strongly affected by stray-light contamination.}
    \label{fig:fullspectrum}
\end{figure*}

\section{The ULX Holmberg II X--1 and its nebula}
\label{sec:sec2}

The object of this study, Ho\,II\,X--1, is located in the  metal-poor dwarf galaxy Holmberg\,II ($Z\,{\sim}\,0.1\,Z_\odot$, SFR\,${\sim}\,0.03\,M_\odot$\, yr$^{-1}$, $d\,{\simeq}\,3.4$\,Mpc; \citealt{Egorov2013, RemyRuyer2015, Karachentsev2003}). 
The ULX is positioned at the heel of the Foot Nebula, an extended, characteristically shaped H\,{\sc ii} region. 
A  more compact He\,{\sc iii} region ($d\sim$100\,pc), discovered by \cite{Pakull2002}, surrounds the X-ray source (see Fig.\,\ref{fig:rgbmap}). 

The {\sl Hubble Space Telescope} ({\em HST}) images showed that the morphology of the He\,{\sc ii}, H$\beta$, and [O\,{\sc i}] emission regions is consistent with the photoionization by the ULX \citep{Kaaret2004}. This is corroborated by optical and UV spectra which are dominated by narrow high-excitation nebular lines, with no evidence for wind-broadening.  Importantly, Ho\,II\,X--1 is one of the few ULXs with a securely identified donor star which has a B-type spectrum and is too cool to significantly contribute to the nebula ionization  \citep{Tao2012,ReyeroSerantes2024}. Therefore, Ho\,II\,X--1 provides a clean room laboratory to study the ionization in the nebula around an accreting black  hole.

In X-rays, Ho\,II\,X--1 is a persistent source with $L_{\rm X}\sim(1-30)\times10^{39}$\,erg\,s$^{-1}$ (0.3--10\,keV) showing substantial variability in luminosity and spectral hardness on multiple timescales \citep{Grise2010,Gurpide2021a}.
No coherent pulsations have been detected \citep{Grise2010,Gurpide2021a}. High-resolution X-ray spectroscopy suggested a tentative evidence for an  absorbing line which may indicate some outflow. The mass of the black hole is estimated as $16 <M_\bullet<70\,M_\odot$ \citep{Barra2024,ReyeroSerantes2024}.

Radio observations resolved the structures within the ULX nebula -- three compact knots (which we name NE, XS, and SW) aligned  along a north-east (NE) -- south-west (SW) axis. The central radio source (XS) is spatially consistent with the X-ray position of the ULX. The steep radio spectrum and subsequent fading of the central source argue against a persistent confined jet, instead favoring  NE and SW radio knots  as intermittent optically thin ejecta associated with past or recurrent jet-related ejecta \citep{Cseh2014,Cseh2015}.

In infrared (IR), \textit{Spitzer} telescope discovered [O\,{\sc iv}] at ${\sim}25.9\,\micron$ in the spectra of Ho\,II\,X--1, however the ULX nebula could not be spatially resolved \citep{Berghea2010a,Berghea2010b}.

\section{Observations and Data Processing}
\label{sec:sec3}

\subsection{Observations and instrument setup}

Ho~II~X--1 was observed on 2025-01-11 19:00:18 UTC with MIRI MRS \citep{Wells2015,Argyriou2023} and on 2025-02-10 at 00:07:43 UTC with NIRSpec   \citep{Boker2022,Jakobsen2022} as a part of the \textit{JWST} program PID\,5846 (PI: L. Oskinova). 

MIRI MRS observations were obtained using four channels, covering the wavelength range from $4.9$ to $27.9\,\micron$ over a FOV up to $6.6''\,\times\,7.7''$ and spectral resolution from $R\sim 3500$ in Channel 1 ($\lambda\,{\simeq}\,5\,\micron$) up to $R\sim 1500$ in Channel 4 ($\lambda\,{\simeq}\,25\,\micron$). 
Each MRS channel was observed in the SHORT and LONG spectral sub-bands, using the SLOWR1 readout pattern with 30 groups per integration and 5 integrations per exposure. A four-point dither pattern was employed, resulting in 20 integrations per sub-band and channel. The total on-source exposure per MRS sub-band and channel resulted in 14716\,s.
In addition, MIRI Imager observations in the F560W filter were obtained in parallel to the MRS exposures, and were used to refine the astrometric calibration. The instrumental set-up was the FASTR1 readout pattern and 30 groups per integration, with a four-point dither pattern.

NIRSpec IFU observations used the G395M/F290LP grating covering  $2.87–5.10\,\micron$ with a resolving power  R${\sim}1000$ over a FOV of $3''\,\times\,3''$.
The data were acquired using the NRSIRS2 readout pattern with 20 groups per integration and a total of 8 integrations with a four-point dither strategy, the total exposure time was 11789\,s. 
No dedicated background exposures were taken for NIRSpec or MIRI.

\subsection{Pipeline calibration}
\label{sec:pipeline}

Raw data are processed through all three stages of the \textit{JWST} calibration pipeline (\texttt{calwebb.detector1}, \texttt{calwebb.spec2}, and \texttt{calwebb.spec3}) using version 1.19.1 with CRDS context \texttt{jwst\_1413.pmap} \citep{Bushouse2022}. Standard detector-level corrections are applied  in Stage~1, including saturation flagging, dark current subtraction, jump detection, and ramp fitting.

For the MIRI\,MRS, Stage~1 cosmic-ray  detection was configured with a lower rejection threshold and extended flagging radii than the defaults to better mask the halos of energetic  events. In Stage~2, we enabled stray-light correction with shower cleaning, a 2D residual fringe correction, and bad pixel self-calibration. In Stage~3, cubes were built per spectral band (\texttt{output\_type}\,$=$\,\texttt{band}) using \texttt{drizzle} weighting, enabling a final band-by-band 1D residual fringe correction; residual bad pixels were replaced using the \texttt{mingrad} algorithm. All {\sc MIRI}\,MRS channels are affected by stray-light contamination, with channels~1 and~2 being the most strongly impacted and consequently the noisiest. To mitigate residual spectral fringing in the {\sc MIRI}\,MRS data, we applied the \texttt{fit\_residual\_fringes\_1d} routine from the {\sc JWST} pipeline prior to spectral extraction.  The origin and characteristics of fringing in the {\sc MIRI}\,MRS  is discussed by \citet{Argyriou2020}.
For the NIRSpec IFU, Stage~1 was configured to detect and expand the flagged area around large cosmic ray snowball events  by a factor of 2. In Stage~3, outlier detection was run with a threshold of  99.7\%. 

No background subtraction is applied. The nebula fills the NIRSpec field of view entirely, and the MIRI\,MRS data are strongly affected by residual stray light that cannot be robustly separated from nebular emission. Consequently, all images, spectra, and emission-line measurements presented in this work should be interpreted as source-plus-background quantities. The local baseline continuum subtraction used commonly for emission-line measurements mitigates the effect of a smooth continuum background, but does not remove any spatially extended line emission present within the field of view. Therefore, absolute line fluxes and line ratios should be interpreted with this limitation in mind, especially for diffuse low-surface-brightness emission, whereas the relative morphology and kinematics of the compact high-ionization lines are less affected.

\subsection{Astrometric calibration}

For the ULX  position we adopt the coordinates of the central (XS) radio source ($\alpha=08^{\rm h}19^{\rm m}28\fs9818 \pm 0\fs0005$, $\delta=+70^\circ42\arcmin18\farcs992 \pm 0\farcs002$), in preference to the \textit{Chandra} X-ray position which has  $0\farcs7$ uncertainty. The NE and SW radio knots are located at separations of ${\sim}$1\arcsec\ and ${\sim}$1.4\arcsec\ from the central source, corresponding to projected distances of ${\sim}$16.5\,pc and ${\sim}$24\,pc respectively. 

The superb blind-pointing accuracy ($\leq$100\,mas) of \textit{JWST} allowed us to avoid target acquisition. The MIRI\,MRS astrometry was refined using a  WCS correction procedure provided by the MIRI instrument team, which registers the MRS data to the \textit{Gaia} DR3 reference frame \citep{Gaia2023} using the simultaneously obtained MIRI\,F560W image. 
For the NIRSpec IFU, we performed a manual astrometric alignment by matching bright H\,{\sc ii} knots above the 99.9th percentile of the \textit{HST}/WFC F502N surface-brightness distribution to the corresponding peaks in the NIRSpec HI\,Br$\alpha$ (${\sim}4.052\,\micron$) map.

\begin{figure*}
    \centering
    \includegraphics[width=0.325\textwidth]{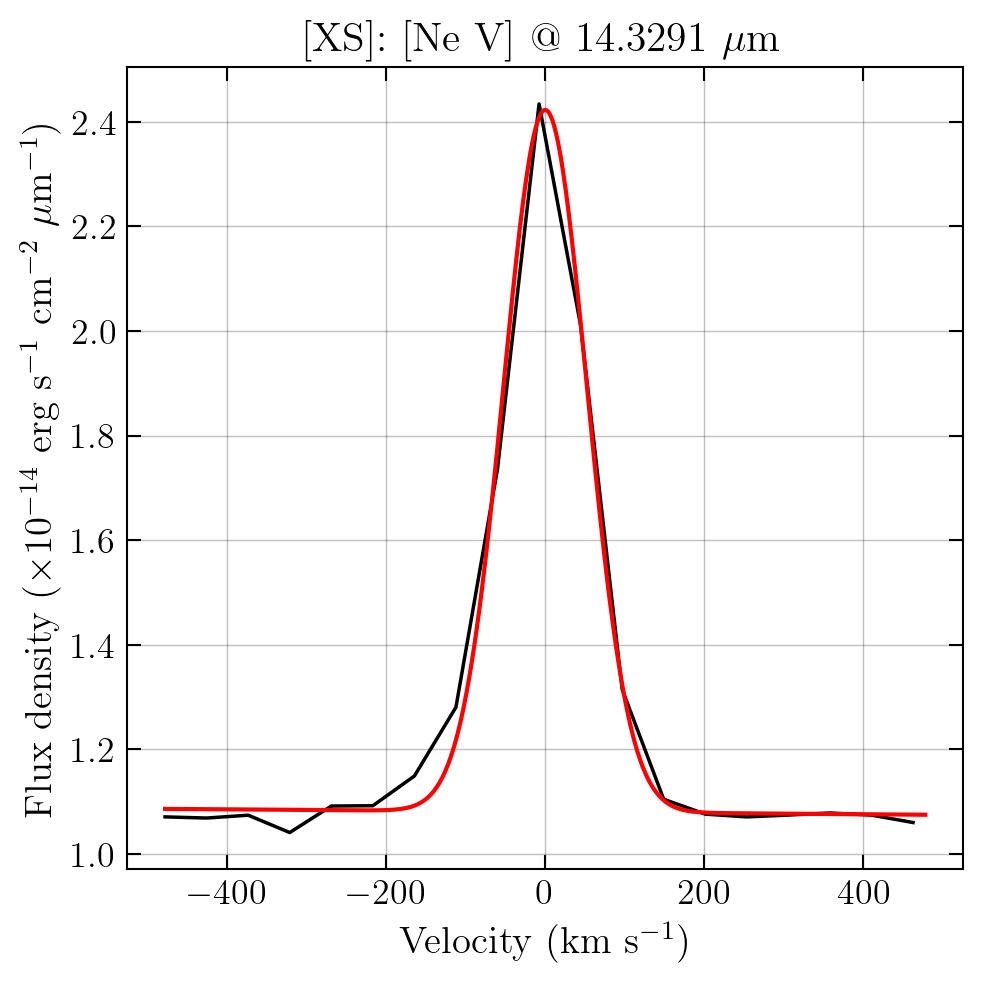}
    \includegraphics[width=0.325\textwidth]{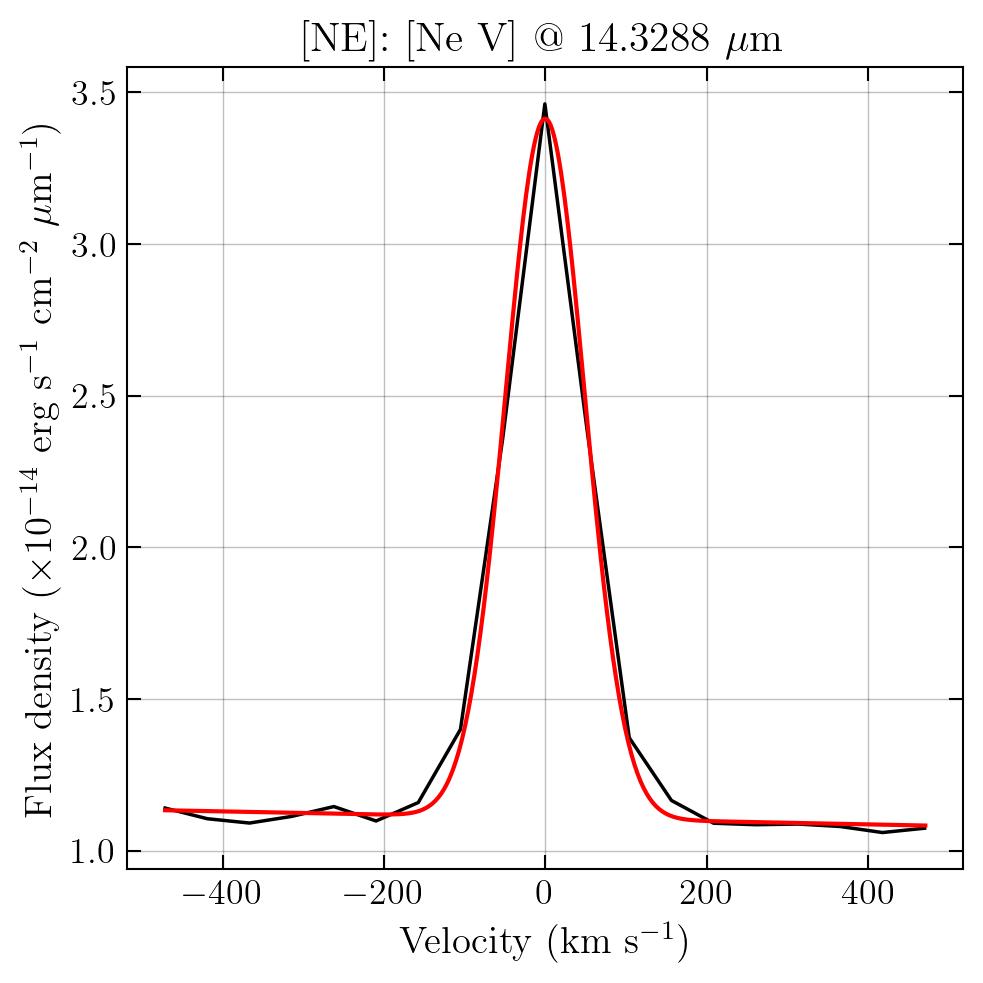}
    \includegraphics[width=0.325\textwidth]{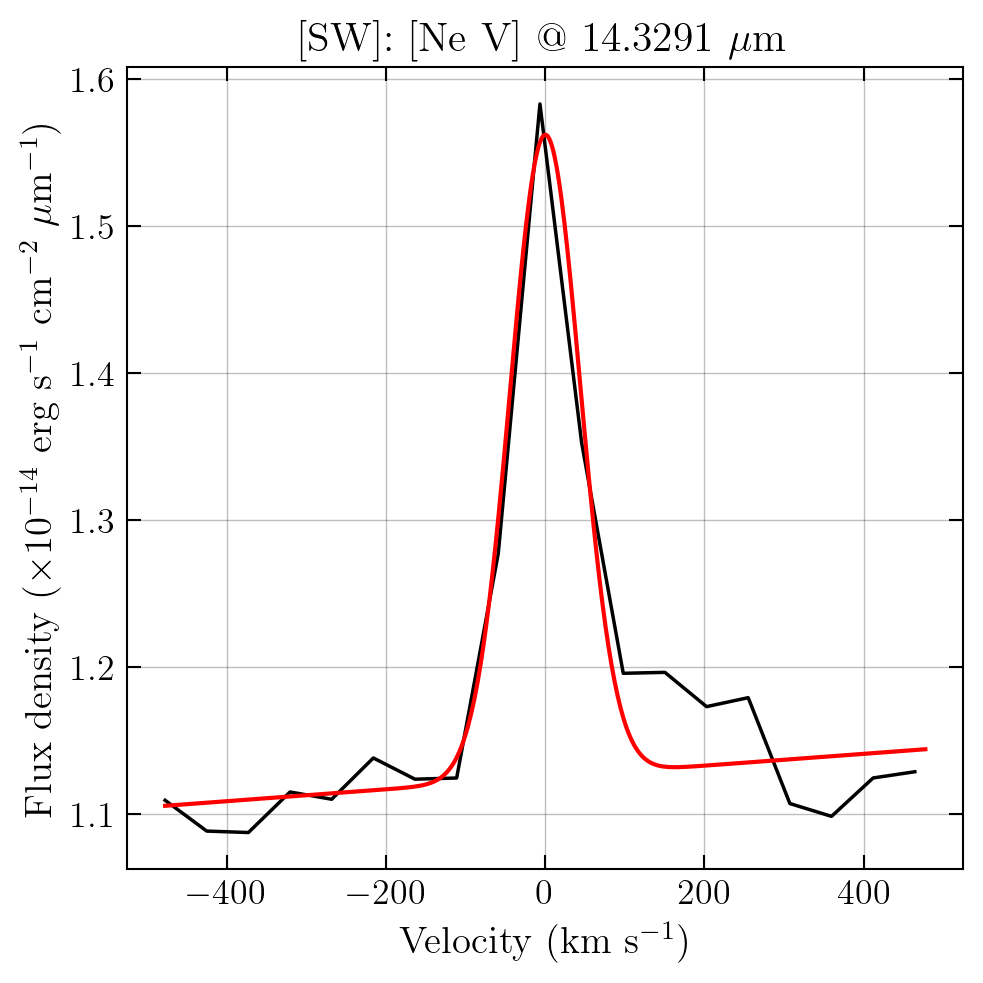}
    \caption{
    Spectra of representative emission lines displayed in the velocity space after correcting for the systemic velocity of the Ho\,II galaxy. Black lines are observations of [Ne\,{\sc v}]${\sim}14.33\,\micron$ extracted from regions XS (left panel), NE (middle panel) and SW (right panel). 
The red curves show the best-fit model consisting of a linear continuum plus a Gaussian profile, with line widths  comparable to the instrumental resolution ($R\,(\lambda\simeq14.3\,\micron)\,{\approx}\,2770$, FWHM$\,\widehat{=}\,110$\,km\,s$^{-1}$). Residual deviations between the data and the best-fit models, most noticeable in the wings of the SW profile, may reflect systematic effects of residual fringing and stray light contamination still present in the data The complete figure set (13 images) is available in the online journal.
    }
    \label{fig:lines}
\end{figure*}

\section{Data Analysis}
\label{sec:sec4}

We assess the overall quality of the calibrated data by extracting a spectrum from a circular region ($d=1\arcsec$) centered at the central source (XS; see Fig.\,\ref{fig:fullspectrum}).
The continuum shows no discontinuities at band or channel boundaries, indicating good cross-calibration both within and between instruments. Several prominent emission lines are clearly seen, including hydrogen recombination lines as well as  numerous fine-structure lines  of ions with $14 < {\rm IP} <126$\,eV.

\subsection{Emission lines }
\label{sec:lineprofiles}

To characterise the kinematic and ionization properties of the nebula, we measure spectra at three positions --- the ULX (XS) and the NE and SW radio knots --- marked by the star and diamond symbols in the right panel of Fig.~\ref{fig:rgbmap}. 
Motivated by the three radio components, at each wavelength slice we model the two-dimensional emission as a linear combination of three PSFs centred at their radio positions.
The three amplitudes are determined simultaneously by weighted least-squares minimization against the per-pixel variance maps \citep{Kamann2013}, yielding deblended one-dimensional spectra. Because the PSF width varies strongly with wavelength, the spatial scale sampled at each position changes across the data set, and Fig.~\ref{fig:rgbmap} therefore marks only the centres of the three components.

For the radio knots, elliptical Gaussian fits to VLA X-band maps give angular sizes of $0\farcs50\times0\farcs30$ (NE) and $0\farcs70\times0\farcs45$ (SW), i.e. ${\sim}8\times5$ and ${\sim}12\times7$\,pc at the distance of Holmberg\,II \citep{Cseh2015}. These scales are comparable to, or smaller than, the JWST PSF over most of the range analysed here: for MIRI\,MRS the FWHM follows $\theta \simeq 0\farcs033\,(\lambda/\mu\mathrm{m}) + 0\farcs106$ \citep{Law2023}, giving $0\farcs36$ at 7.65\,\micron\ ([Ne\,{\sc vi}]), $0\farcs58$ at 14.32\,\micron\ ([Ne\,{\sc v}]) and $0\farcs96$ at 25.89\,\micron\ ([O\,{\sc iv}]).  
The point-source approximation is therefore most appropriate at the longest MIRI wavelengths, becomes less accurate around [Ne\,{\sc v}], and is least justified for [Ne\,{\sc vi}] and the NIRSpec lines. This approximation may affect the recovered fluxes of spatially extended components, but does not affect the surface-brightness maps and the centroid/opening-angle measurements presented in Sec.\,\ref{sec:sbmaps}

Lines are detected in the continuum-subtracted spectra using a signal-to-noise--based algorithm, adopting ${\rm SNR}\,{\ge}\,30$ to account for the elevated and spatially variable noise in the MIRI data caused by straylight.
For each detected feature, a Gaussian profile plus a linear baseline was fitted in velocity space within a velocity window of ${\pm}\,500\,{\rm km\,s^{-1}}$ for MIRI MRS and $\pm\,800$\,km\,s$^{-1}$ for NIRSpec. The best-fitting models are used to measure the central wavelength, velocity shift, velocity dispersion, and integrated line flux, as well as 68\% confidence-level uncertainties.
The observed wavelengths are corrected for the systemic velocity of the Holmberg\,II galaxy ($+\,157$\,km\,s$^{-1}$; \citealp{Egorov2013}). 
The lines are identified using the Atomic Line List~v3.00b5 \citep{vanHoof2018}, restricting the search to transitions expected in ionized nebular gas and prioritizing elements seen in other ULX nebulae in low metallicity galaxies like I\,Zw\,18  and SBS\,0335\,--\,052E \citep{Hunt2025a,Mingozzi2025}.

Representative  recombination and forbidden lines detected in the XS, NE and SW regions are shown in Figs.\,\ref{fig:lines}. All  lines have symmetric profiles with widths comparable to the instrumental spectral resolution. No evidence for broad wings,  or multiple kinematic components is detected in any of the lines, including the highest-ionization species. The line centroids are consistent with zero velocity within the uncertainties. These profiles indicate that the nebula gas traced by both low- and high-ionization species is kinematically quiescent. 

\begin{table}
\centering
\small \begin{tabular}{@{}lccccc@{}}
\toprule
Line & $\lambda_{\rm rest}$ & IP & $\Delta r$ & PA & OA \\
 & ($\mu$m) & (eV) & (\arcsec) & (deg) & (deg) \\
\midrule
{H\,{\sc i} Br$\alpha$} & 4.052  & 13.6  & $0.41^{+0.13}_{-0.11}$ & $-46^{+15}_{-14}$ & $136^{+23}_{-32}$ \\
{[Ne\,{\sc ii}]}        & 12.814 & 21.6  & $0.67\pm0.10$          & $13^{+17}_{-9}$   & $80^{+16}_{-18}$  \\
{[S\,{\sc iii}]}        & 18.713 & 23.3  & $0.55^{+0.10}_{-0.08}$ & $24^{+14}_{-9}$   & $95^{+15}_{-20}$  \\
{[Ar\,{\sc iii}]}       & 8.991  & 27.6  & $0.38^{+0.12}_{-0.10}$ & $38^{+16}_{-15}$  & $93^{+35}_{-28}$  \\
{[S\,{\sc iv}]}         & 10.511 & 34.8  & $0.45^{+0.12}_{-0.08}$ & $54^{+14}_{-15}$  & $68\pm13$         \\
{[Ne\,{\sc iii}]}       & 15.556 & 41.0  & $0.75^{+0.10}_{-0.09}$ & $11^{+16}_{-8}$   & $68^{+14}_{-15}$  \\
{[O\,{\sc iv}]}         & 25.890 & 54.9  & $0.82^{+0.13}_{-0.12}$ & $49^{+6}_{-8}$    & $48^{+12}_{-11}$  \\
{[Ar\,{\sc vi}]}        & 4.530  & 75.0  & $1.11^{+0.12}_{-0.12}$ & $53^{+5}_{-5}$    & $39^{+6}_{-7}$    \\
{[Ne\,{\sc v}]}         & 14.322 & 97.1  & $1.16^{+0.11}_{-0.10}$ & $68^{+6}_{-4}$    & $38^{+8}_{-7}$    \\
{[Ne\,{\sc v}]}         & 24.318 & 97.1  & $1.19^{+0.18}_{-0.16}$ & $63^{+9}_{-18}$   & $57^{+38}_{-17}$  \\
{[Ne\,{\sc vi}]}        & 7.652  & 126.2 & $1.46^{+0.09}_{-0.11}$ & $68^{+5}_{-4}$    & $34^{+7}_{-8}$    \\
\bottomrule
\end{tabular}\caption{Areas in the brightness maps filled by line emission ($1''\,{\simeq}\,16.5\,{\rm pc}$). Separation $\Delta r$ and position angles (PA) are measured from the ULX position (XS) to the flux-weighted centroid computed within the 68\% surface-brightness contour of the line. Position angles (PA) are given in degrees east of north. Opening angles (OA) are measured in the direction perpendicular to the axis defined by the line connecting XS and the flux-weighted centroid. Uncertainties correspond to the 68\% confidence interval derived from $10^3$ Monte Carlo realizations, accounting for flux uncertainties, astrometric errors, and variations in the adopted contour level.}
\label{tab:line_geometry_by_ip_ascending}
\end{table}

\begin{figure*}
    \centering
    \includegraphics[width=0.49\textwidth]{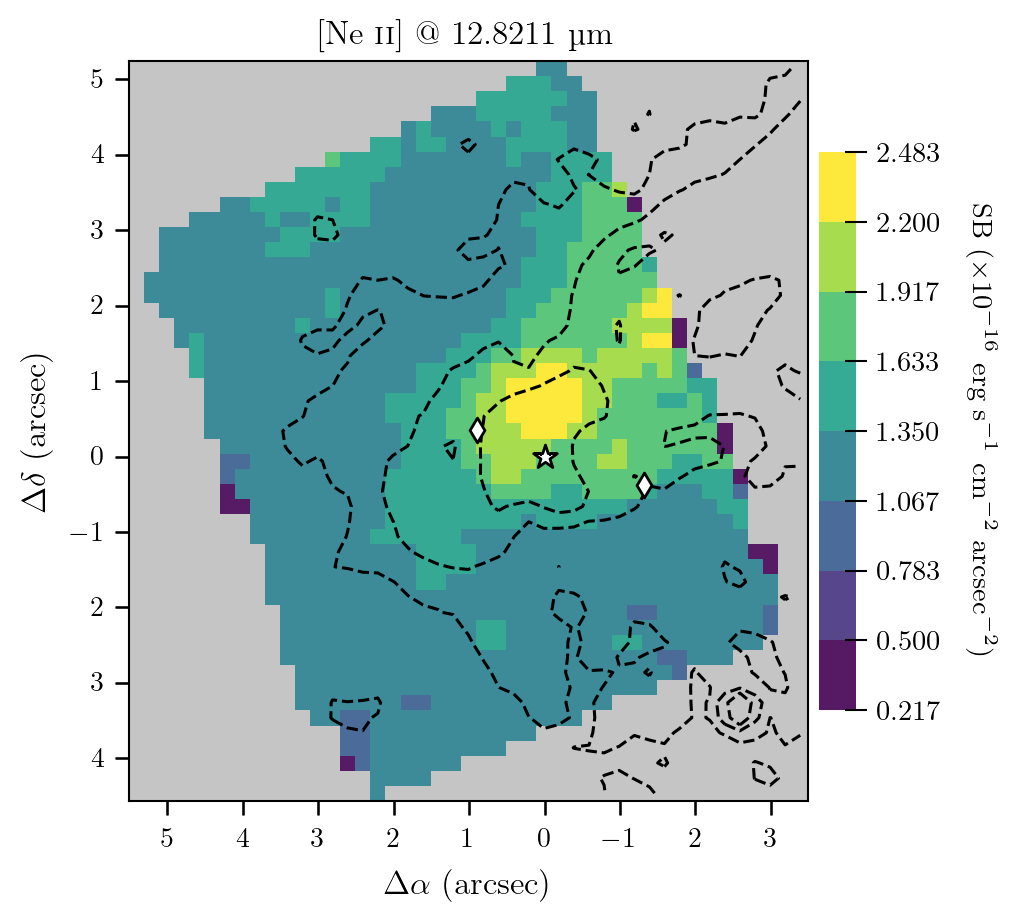}
    \includegraphics[width=0.49\textwidth]{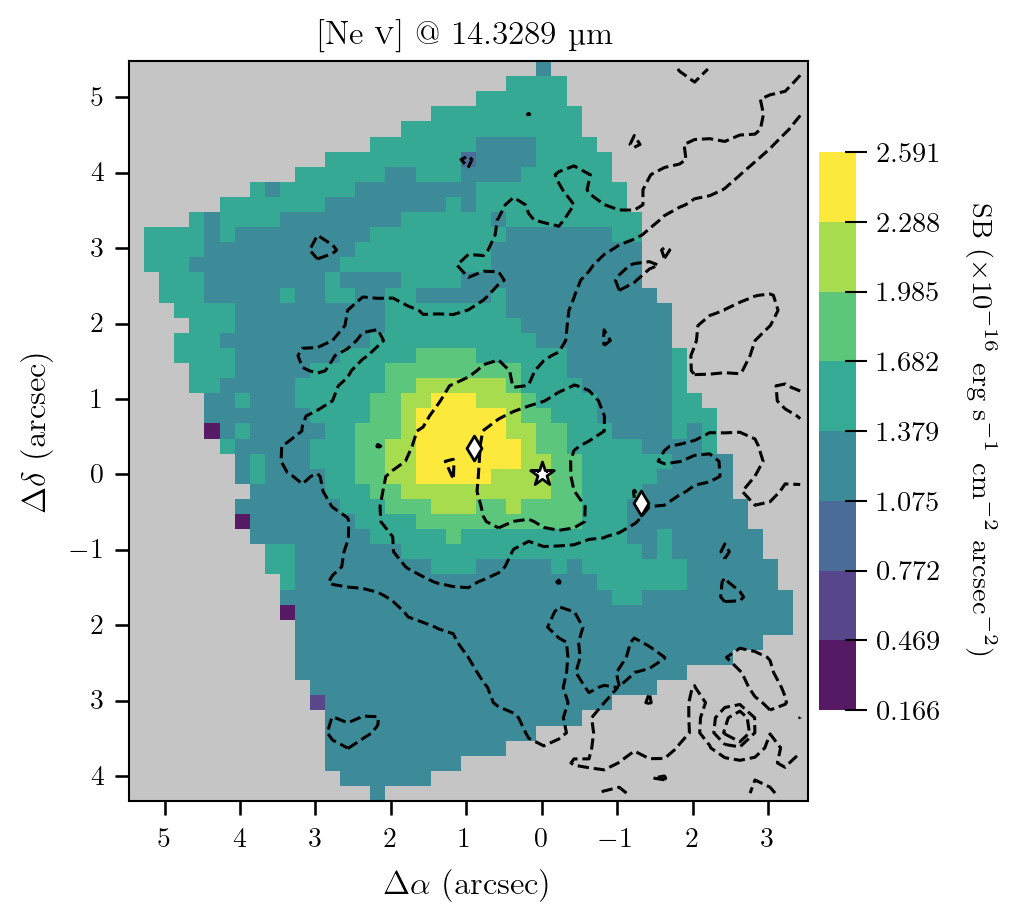}    
    \caption{
    Line surface brightness maps of [Ne\,{\sc ii}] (IP$\,{\sim}\,21.6$\,eV) and [Ne\,{\sc v}] (IP$\,{\sim}\,97.1$\,eV) detected in the MIRI MRS data of Ho\,II\,X--1 nebula ($1\,\arcsec\,{\simeq}\,16.5$\,pc) expressed in celestial offset coordinates with respect to the ULX position (XS, white star symbol). The color scale shows the line-integrated surface brightness in units of erg\,s$^{-1}$\,cm$^{-2}$\,arcsec$^{-2}$. Only spaxels with SNR\,${\geq}\,3$ are shown.  Black dashed contours show the 95th, 99th, and 99.7th percentile levels of the HST/ACS F502N surface brightness emission of Ho\,II\,X--1 smoothed with a Gaussian filter of 0\farcs25 width. The diamond markers indicate the positions of the NE and SW radio knots. The complete figure set (11 images) is available in the online journal.
    }
    \label{fig:maps}
\end{figure*}

\subsection{Line surface-brightness maps}
\label{sec:sbmaps}

We construct two-dimensional maps of line surface brightness from the calibrated spectral cubes. The maps are derived on a spaxel-by-spaxel basis using local continuum subtraction and direct spectral integration.
For each spaxel, a continuum is fitted using wavelength sidebands adjacent to the line. Line surface brightness is obtained by integrating the continuum-subtracted spectrum within a window of $\pm\,3\sigma_{\lambda}$ around the  centroid wavelength, with the continuum estimated from sidebands spanning $5\sigma_{\lambda}$\,--\,$8\sigma_{\lambda}$ on either side of the feature. Only spaxels with positive surface brightness and $\mathrm{SNR}\,{\ge}\,3$ are kept. 

 In Figs.\,\ref{fig:maps} we present surface-brightness maps of representative lower- and higher-ionization species of [Ne\,{\sc ii}] $\lambda 12.82\,\micron$ and [Ne\,{\sc v}] $\lambda 14.33\,\micron$. The maps clearly show the structures and stratification in the spatial distribution of the ionized gas within the nebula.

The areas traced by hydrogen lines and lines of low-ionization species, such as [Ne\,{\sc ii}] and [S\,{\sc iii}], are extended and roughly follow the optical nebular structures traced by the {\em HST} images (e.g.\,[O\,{\sc iii}]\,$\lambda 5007$\,\AA\ contours). 
In contrast, the areas traced by lines of highly ionized species  including [Ne\,{\sc v}], [O\,{\sc iv}], and [Ne\,{\sc vi}], are significantly more compact. Moreover, the emission is not distributed symmetrically around the ULX position, but has a preferential extension toward the north-east, with a secondary dimmer extension toward the south-west.

To quantify these trends, we measure for each transition the flux-weighted centroid within the 68\% surface-brightness contour and compute its projected separation ($\Delta\,r$), position angle (PA), and opening angle (OA) relative to the ULX position (Table\,\ref{tab:line_geometry_by_ip_ascending}).
Areas filled with the line-emission from the highly ionized species with
IP\,${\gtrsim}$\,97\,eV show systematically smaller opening angles and a remarkable
coherence in position angle, clustering around $\mathrm{PA}\,{\sim}\,68^\circ$. In
particular, bright areas filled with [Ne\,{\sc v}] and [Ne\,{\sc vi}] line
emission are offset by $17$\,--\,$24$\,pc from the ULX. This shows that the
highest-ionization gas is compact while the lower-ionization material has a more
extended and quasi-isotropic distribution. This behaviour is not an artefact of the
wavelength-dependent angular resolution: H\,{\sc i}\,Br\,$\alpha$ (4.05\,\micron,
$\mathrm{IP}=13.6$\,eV) and [Ar\,{\sc vi}] (4.53\,\micron, $\mathrm{IP}=75$\,eV),
observed with essentially the same PSF, differ by a factor of ${\sim}3.5$ in opening
angle ($136^{\circ}$ versus $39^{\circ}$).

\section{Discussion}
\label{sec:sec5}

Our JWST observations confirm the [O\,{\sc iv}] emission previously detected with Spitzer \citep{Berghea2010a,Berghea2010b} and reveal [Ne\,{\sc v}] and [Ne\,{\sc vi}] for the first time in the nebula around Ho II X-1. \textit{JWST} mapped the spatial distribution of ionized gas on projected spatial scales of ${\sim}15-25$\,pc. We find a clear ionization-dependent morphology: lower-ionization emission extends over a broad region around the ULX, whereas the highest-ionization species are more compact and displaced from the central source.

In models of supercritical accretion, an optically thick disk wind forms a funnel that preferentially channels radiation from the inner accretion flow along the polar direction \citep{Poutanen2007,King2009}. Our observations indicate that the angular confinement of the nebular emission increases with ionization potential. The opening angle decreases from ${\sim}136^\circ$ for H\,{\sc i}\,Br$\alpha$ to ${\sim}38^\circ$ for [Ne\,{\sc v}] and ${\sim}34^\circ$ for [Ne\,{\sc vi}], with the highest-ionization species sharing a position angle of ${\sim}68^\circ$. This pattern is consistent with the highest-energy ionizing photons escaping preferentially along the polar direction. Conversely, the broader He\,{\sc ii} morphology previously resolved with HST indicates that lower-energy ionizing radiation is distributed over a wider solid angle \citep{Kaaret2004}. Ionization cones around stellar-mass black holes are rare; LMC X-1 provides the clearest comparison, where the ionized morphology also only roughly approximates an ideal cone \citep{Cooke2008}.

The highest-ionization emission is also aligned with the NE--SW radio knots, suggesting a physical connection between the jet axis and the directions along which the hardest ionizing radiation reaches the nebula. The [Ne\,{\sc v}] and [Ne\,{\sc vi}] emitting regions are displaced by $17-24$\,pc from the ULX, corresponding to a minimum light-travel time of several decades. The nebular morphology therefore probes the past ionizing radiation field rather than the present X-ray spectral regime or instantaneous viewing geometry. One possibility is that past jet activity modified the surrounding medium along the same axis, facilitating the propagation of ionizing radiation. Pre-existing density inhomogeneities in the ambient ISM may additionally shape the detailed morphology of the ionized emission.

The nebular gas is kinematically quiescent across all ionization stages probed by our data, with line widths consistent with the instrumental resolution and no evidence for fast bulk motions. This contrasts with shock-dominated ULX bubbles, where optical emission lines commonly display expansion velocities of several hundred km\,s$^{-1}$ and asymmetric profiles \citep{Pakull2002,Gurpide2022,Gong2023}. If fast radiative shocks were responsible for producing [Ne\,{\sc v}] and [Ne\,{\sc vi}] in Ho\,II\,X-1, substantially higher velocities would be required \citep{Dopita1996,Allen2008}. The absence of such signatures favors photoionization by the ULX as the dominant excitation mechanism, consistent with the absence of strong, fast disk-wind signatures in the UV and optical spectra analyzed by \citet{ReyeroSerantes2024}.

\section{Conclusions}
\label{sec:sec6}

\begin{enumerate}
    \itemsep0.05em

    \item The radiation of the ULX  Ho\,II\,X--1 is responsible for the extreme ionization conditions in the surrounding nebula. All detected emission lines exhibit narrow widths with no evidence for fast gas bulk motions. These results disfavor fast radiative shocks as the dominant excitation mechanism. 
    
    \item The ionized gas exhibits a clear morphological stratification with ionization potential. Hydrogen lines and low-ionization species trace the extended nebular structure. On the other hand, [O\,{\sc iv}], [Ne\,{\sc v}], and [Ne\,{\sc vi}] emission regions are confined and are aligned with the NE--SW axis traced by radio knots. This morphology is consistent with anisotropic escape of the soft-X-ray ionizing continuum  and suggests that the angular dependence of ULX emission may extend below the hard X-ray regime.

    \item The spatial coincidence between the highly ionized gas and the radio knots, and the $>65$ light-years distance between the central engine and the cloud of most highly ionized gas, suggests a physical connection between past or recurrent jet-ejection activity and the paths along which the ionizing radiation reaches the nebula.

    \item The \textit{JWST} observations have demonstrated that an ULX powered by an accreting black hole in a metal poor galaxy is responsible for the extreme photoionization of the surrounding gas which  emits a spectrum dominated by the lines of ionic species with IP$> 54$\,eV.

\end{enumerate}

\begin{acknowledgments}
The authors are grateful to Daniel Schaerer for useful comments and suggestions.
The authors are grateful to the MIRI calibration team for their support with the MIRI MRS WCS alignment corrections.
This work is based on observations made with the NASA/ESA/CSA James Webb Space Telescope. These observations are associated with program \#5846. The data were obtained from the Mikulski Archive for Space Telescopes at the Space Telescope Science Institute (\dataset[10.17909/q0pd-hc75]{http://dx.doi.org/10.17909/q0pd-hc75}), which is operated by the Association of Universities for Research in Astronomy, Inc., under NASA contract NAS 5-03127 for JWST. 
TB acknowledges support by the German Aerospace Center (DLR) grant 50OR2603. 
DM acknowledges support by the grant PID2024-155316NB-I00 funded by MICIU/AEI/10.13039/501100011033/FEDER, UE and CSIC PIE 202350E189, as well as the Spanish program Unidad de Excelencia María de Maeztu financed by MCIN/AEI/10.13039/501100011033, and by the MaX-CSIC Excellence Award MaX4-SOMMA-ICE. 
DP acknowledges financial support from the FWO in the form of a junior postdoctoral fellowship No. 1256225N.
\end{acknowledgments}

\facilities{JWST (NIRSpec, MIRI), HST (WFC)}
\software{Astropy \citep{astropy2013,astropy2018,astropy2022}, NumPy \citep{NumPy2020}, SciPy \citep{SciPy2020}, STPSF \citep{Perrin2012,Perrin2014}, Specutils \citep{Specutils} }

\clearpage
\bibliography{new.ms}{}
\bibliographystyle{aasjournalv7}

\end{document}